\documentclass{article}

{}
{}
{}

\usepackage{amsmath} 
\usepackage{amsthm} 
\usepackage{cite} 
\usepackage{graphicx}
\usepackage{algorithmic} 
\usepackage{authblk}
\usepackage[margin=1.2cm]{geometry}
\usepackage{multicol}
\usepackage{blindtext}
\usepackage{epstopdf}

\usepackage[hyphens]{url}
\usepackage[hyphenbreaks]{breakurl}

\begin{document}

\title{\textbf{Onboarding Citizens to Digital Identity Systems}}

\author[1*]{Tasos Spiliotopoulos}
\author[2]{Al Tariq Sheik}
\author[3]{Debora Gottardello}
\author[4]{Robert Dover}

\affil[1]{{School of Computer Science, University of Birmingham, Birmingham, UK}}
\affil[2]{{The Alan Turing Institute, London, UK}}
\affil[3]{{Edinburgh Business School, University of Edinburgh, Edinburgh, UK}}
\affil[4]{{School of Criminology, Sociology and Policing, University of Hull, Hull, UK}}
\affil[*]{{Email: a.spiliotopoulos@bham.ac.uk}}

\date{}

\setcounter{Maxaffil}{0}
\renewcommand\Affilfont{\itshape\small}

\maketitle

\begin{abstract}
Digital Identity (DI) technologies have the potential to enhance the quality of life of citizens through the provision of seamless services, improve the effectiveness of public services, and increase overall economic competitiveness. However, lack of access to DIs can limit these benefits, while unequal access can lead to uneven distribution of these benefits across social groups and escalate existing tensions. Accessible, user-friendly and efficient onboarding can play a key role in ensuring equitable access and wide adoption of DI technologies. This paper proposes the development of physical locations (Experience Centres) that can be used for citizen onboarding to national DI systems, positively shaping citizens' first impression with the technology and, in turn, promoting adoption. To this end, we outline a multidisciplinary research approach for identifying and addressing the considerations necessary for designing, developing and operating a model Experience Centre for DI onboarding in an inclusive manner.
\end{abstract}

\begin{multicols}{2}

\section{Introduction}

In recent years, policymakers, researchers, and practitioners around the globe have recognised the potential benefits of Digital Identity (DI) systems \cite{Tan2022}. Governments have begun implementing digital identity programmes to provide legal and regulated digital identities to citizens. The UK government has taken a large step in this direction by publishing a beta version of the \textit{Digital Identity and Attributes Trust Framework} \cite{UKGovernment2023}, which is intended to provide a policy framework  that enables and encourages the ability for individuals to have reusable certified Digital IDs.
\vskip1pc
DIs provide citizens with easy, efficient, privacy-preserving and secure access to services. This, in turn, allows governments and businesses to innovate, streamline their services, comply with regulations and compete at the international level. For example, researchers have highlighted how the selective disclosure and self sovereignty afforded by DI technologies can address financial exclusion \cite{Spiliotopoulos2021, Wang2020}, while also supporting innovation \cite{Elliott2022}.
\vskip1pc
Despite the potential advantages, research indicates that the design and implementation of national DI systems may have significant socio-economic, ethical, privacy, and human rights implications \cite{Beduschi2021}. When designing and implementing DI systems, these implications must be carefully considered, as they have the potential to affect a wide variety of individuals and groups. One important consideration involves the onboarding process, which needs to be designed in a way that promotes equitable access to DIs for all citizens. This paper outlines a research approach for identifying and addressing the considerations necessary for designing, developing and operating a physical location (an Experience Centre) for DI onboarding in an inclusive manner.

\section{Digital Identity Technologies}
\subsection{Technical Background}

In broad terms, a digital identity refers to what an entity, object or subject is \cite{Ante2022}. A key characteristic of a digital identity is that it can be used to prove something about an entity. This means that a third entity can verify a claim that an issuer of the identity has made about the identity holder. This, in turn, means that services and organisations can trust these claims. With this in mind, Kameron \cite{Kameron2005} has defined a digital identity as 'a set of claims made by one digital subject about itself or another digital subject'. When considering identities at the national level, the UK Digital Identity and Attributes Trust Framework refers to a DI as a 'digital representation of a person acting as an individual or as a representative of an organisation' and highlights the importance of the ability for people to prove claims about themselves and the impact that this foundation of trust can have on organisations, service providers and the country's economy \cite{UKGovernment2023}.
\vskip1pc
A DI \textit{system} is a mechanism that permits the creation and verification of an individual's identity using digital means. The process of using an identity within this system consists primarily of two steps: (i) onboarding, and (ii) authentication and ID management. During the initial phase of the onboarding procedure, an individual's Personal Identifiable Information (PII) is collected, validated and verified. This information is used to identify and establish the user within the system. Document validation, email verification, and phone verification may be included in the de-duplication and verification process. Once the individual's identity has been proven, the administrator facilitates in creating an identity record. The second stage of the process is authentication and ID management, in which the individual's identity is verified and managed whenever they attempt to use a service. This is achieved through a variety of methods, including password-based authentication, two-factor authentication, biometric authentication, issuing, recording credential, binding, expiration, renewal and revocation. The authentication phase's objective is to ensure that only authorised users can access the service and their credentials are managed.
\subsection{Policy Background}

The UK Digital Identity and Attributes Trust Framework (DIATF) provides a preliminary and evolving set of rules and standards for those providing (or independently certifying) DIs. The DI providers will need to follow these rules and standards in order to provide secure and trustworthy digital identity and attribute solutions in the UK market. The drafting of the DIATF is being overseen by the UK government's Department for Digital, Culture, Media and Sport (DCMS), but this process involves a number of government, scientific, policy and corporate stakeholders injecting expertise into the process. At the time of writing there is no definitive publication date for the final version of the DIATF, with the DCMS stating that it will be published in the short term. However, the currently published beta version makes clear the government's intention to use the trust framework to enable services such as digital right to work, rent and criminal record checks \cite{UKGovernment2023}. 
\vskip1pc
The DIATF is an important element of the UK government's digital economy initiatives that they argue will facilitate increased levels of innovation, competition, and transparency into the digital identity market. The initial DIATF also puts special emphasis on the protection of individual privacy and security. Such ambitions are tempered by the government’s parallel policy agenda of deanonymising individuals to improve the work and effectiveness of law enforcement and intelligence agencies.
\vskip1pc
The transnationality of digital markets sits uneasily with Westphalian constructs of sovereignty, and so the DIATF will need to be compatible (in important ways) with the European Union’s proposed European Digital Identity Framework \cite{Commission2021}, which builds on the existing cross-border legal framework for trusted digital identities, the European electronic identification and trust services initiative (eIDAS Regulation) \cite{Commission2022}.

\subsection{Considerations around Access to DI Technologies}

DI systems, when not designed appropriately, may fail to address the needs of those who already carry significant markers of social and economic marginalisation. Such exclusion results in negative economic consequences and serves to further exclude them from formal mechanisms based upon trustworthy identity systems. For example, biometric identification methods, such as fingerprints, may not be accessible to disabled or elderly individuals, creating 'digital barriers' to access economic and social resources \cite{Masiero2021, Martin2021}. Additionally, digital identification may enable more efficient discrimination against marginalised groups, such as women, ethnic minorities, religious groups, disabled individuals, and members of the LGBT community \cite{Davis2019}. These systems also pose a threat to personal safety of marginalised groups. Therefore, more trustworthy digital identity systems that address the rights of all individuals, particularly the most marginalised, will be necessary in the future.
\vskip1pc
Research has shown that in the UK, around 11 million people especially from marginalised backgrounds do not have a passport or a driving licence. Women, those living in urban areas, the under 20s and over 65s are less likely to hold a driving licence \cite{Commission2019}. Data from the UK Electoral Commission reported that disadvantaged groups are more likely to not have an ID. For example, older people (aged 85+) were less likely to have an ID that was recognisable (91\% compared to 95\%–98\% for those in younger age groups. It also found that the unemployed, people with disabilities, and people without qualifications, were all less likely to hold any form of photo ID \cite{Society2021}. 
\vskip1pc
Onboarding can be difficult for people who do not have access to technological resources or who are not technologically skilled. Not having an ID can have important implications for marginalised people as they may lack access to services that ensure credit accumulation or profit storage. As a consequence, marginalised people may have difficulty accessing many basic services, including work, social protection, banking or education. Likewise, the lack of a documented identity puts vulnerable and already marginalised people at constant risk of transgressing the lines between legal and illegal.
\vskip1pc
In recent years, digital identity verification through a mobile account has proven to be an effective verification method in several countries. This method has been used to prove identity in order to receive benefits from the government, private entities, or obtain microloans. Accessing financial services online or through mobile devices provides independence, the opportunity to pay daily expenses and to make longer-term plans, and therefore remove a key source of anxiety \cite{Bruggen2017}. Having access to financial services helps marginalised groups not only to survive but also to bring themselves closer to the mainstream of society in terms of access whilst maintaining their individual identities, potentially facilitating a greater level of respect from those who do not find themselves marginalised. Poverty or social isolation driven by lack of access to services, including financial services, affects minorities in all nations. 
\vskip1pc
In the western world, communities who are seen as being marginalised often correlate with low recorded levels of literacy, a lack of access to financial services and consequently a reliance on outside agencies, making this loosely confederated group a challenge to onboard to digital services \cite{Ozili2021}. Despite the fact that the richer countries, such as the United Kingdom, tend to have better quality services than poorer nations, the security that prevents unauthorised use of these services is much stricter. The cost of buying a passport or learning to drive to obtain a photo ID can easily prevent minorities from possessing these essential ‘entrance’ documents.

\section{Digital Identity Onboarding}

As noted above, a DI system is a mechanism that permits the creation and verification of an individual's identity using digital means. The process of using an identity within this system consists primarily of two steps: (i) onboarding, and (ii) authentication and ID management.
\vskip1pc
The implementation of these two phases incurs respective challenges. The onboarding phase contains many challenges that are social, economic, political and technological in nature. These include data collection, data verification, privacy, security, user experience, scalability and compliance. 
\vskip1pc
The challenges faced during the authentication phase and ID management are analogous to those experienced in technological developments, for example: security, usability, scalability, false acceptance rate and false rejection rate, privacy, compliance and interoperability. Unlike the technological challenges in the authentication phase, which are dependent on the initial stage, the challenges in the onboarding phase are interdisciplinary in nature, making onboarding the referent object for a multidisciplinary inquiry, and requires a number of disciplinary perspectives and injects to generate solutions. As a result, it is crucial to formally address the challenges of the onboarding phase and design interdisciplinary solutions to create a trustable, efficient, and user-friendly experience.
\vskip1pc
The main obstacles in adopting DIs are: 1) the information gap that exists between the consuming public and the technology companies, and 2)  people’s hesitation to initially engage with the technology. Trust in technology in general, trust in a specific technology, and trust in the people and institutions behind a technology play an important role in shaping people’s beliefs and behaviour \cite{Mcknight2011, Guggenberger2023}. To establish trust, the DI system’s onboarding, authentication, and ID lifecycle management processes must be demonstrated as trustworthy: this is both a measure that can be technically benchmarked and is also subject to sentiment. The consuming public's first impression and initial experience with a technology are also particularly important in shaping adoption and post-adoption behaviours \cite{Fox2021}. Because these early beliefs and behaviours establish a path-dependency, we are identifying the DI onboarding phase as a key research consideration for ensuring equitable access and wide adoption of digital identities.

\section{Achieving a Smooth Digital Identity Onboarding Experience}

In order to increase the adoption of DIs, and to do so in a fair and equitable manner, we propose the use of physical locations for citizen onboarding to DI systems in the UK context. Such an approach has similarities to the use of Experience Centres (ECs) developed in other countries\footnote{https://mosip.io/news-events/announcing-the-launch-of-the-first-mosip-experience-centre-an-end-to-end-walk-in-mosip-experience-in-bangalore-india}. These ECs, which are physical locations, will allow users to register and collect digital IDs and credentials, and integrate them with other systems and services such as civil registration systems, e-sign, and electronic health records management. 
\vskip1pc
An Experience Centre can facilitate trustworthy and inclusive onboarding to DI technologies. This has the potential both to address uneven access to DI technologies, and to increase DI adoption overall. Ensuring that access to DI technologies is inclusive can profoundly reduce inequalities as proving one's identity is rapidly becoming an essential part of exercising human rights on a day-to-day basis. The use of an EC is also to improve the efficiency of the DI onboarding process with additional services taking place on-site, such as document verification, biometric capture, and identity document scanning. Such an EC would provide a secure and user-friendly environment for users to interact with the DI system, close the information gap between the public and DI providers, and develop confidence in the technology and related services. ECs also allow us to iterate and improve the user experience within them, thus constantly improving accessibility and trust. We envision an EC as a \textit{Digital Identity Playground} that can positively shape citizens' first impression with the technology and, in turn, promote adoption of DI technologies.
\vskip1pc
ECs are complex sociotechnical systems and, as such, are very difficult to design and implement \cite{Baxter2011}. A number of considerations need to be taken into account in order to address fundamental design questions, such as:
\begin{itemize}
    \item What are the most important features and services of a model EC in the UK?
    \item What are the specific requirements of a model EC in terms of technical infrastructure,
staffing, spatial architecture, cost and security?
    \item What are the main design considerations to ensure that the model EC can engage citizens in an inclusive manner, increase adoption, build confidence in the use of DIs, and act effectively as a digital playground?
    
\end{itemize}

\section{Our Research Approach}

To operationalise our \textit{multidisciplinary} approach we have designed a series of research activities and methods that are necessary steps to effectively address these questions. The contribution from Foreign Policy Analysis (FPA) of \textit{Horizon scanning} will be used to identify the key drivers shaping the DI onboarding operational environment and key action points to proactively shape desirable futures. The output of a horizon scan is a formal assessment document that provides a probabilistic measure of likelihood of various future trends occurring and allows the recipient to make evidence based judgements about resourcing and framing responses to the initial challenge. In this context, the horizon scan will identify trends over the ten-year time period, from most likely to wild card possibilities, and also provide assessments of the sourcing base for these judgements. The output from the horizon scan will exist as a standalone document, but also helps to inform the creation of requirements (e.g., specific use cases and design diagrams) and recommendations, that in turn provide an underpinning for the design of an Experience Centre. In general terms, a horizon scan is an empiricist tool for identifying the key elements of the phenomena or issue in hand - in this case onboarding. Further, a horizon scan assists in generating areas for further research, action and mitigation \cite{Hines2019}. 
\vskip1pc
The contribution from Operational Sciences and Human-Computer Interaction is a \textit{Literature review and science mapping analysis} which aims at investigating the state of the current research and also the implementation trends and opportunities in DIs with a specific focus on the onboarding process. This structured analysis of this large body of academic information relating to DIs will allow us to infer research trends over time, recognise themes, identify shifts in the boundaries of the disciplines, detect the most prolific scholars, institutions and countries, and to present the 'big picture' of extant research around DI systems, user adoption and onboarding \cite{Chen2017, Aria2017}.
\vskip1pc
\textit{Semi-structured interviews} - derived from a social scientific underpinning - will help to provide an understanding of stakeholder and end-user perceptions and attitudes towards DI systems, with a focus on inclusivity as a framing device within onboarding. The main purpose of the interviews with stakeholders will be to investigate possible challenges, barriers, attitudes and opportunities and identify major trends to inform the development and design of future trustworthy digital identities that guarantee equality and inclusion and are accessible for all. Moreover, by interviewing stakeholders we will be able to understand how ECs can ensure an equal and inclusive society. 
\vskip1pc
Finally, \textit{Threat and risk assessment}  will determine the methods, practices, and approaches that provide the greatest traction for identifying and assessing security threats and evaluating the associated risks for inclusivity in future digital identity onboarding systems. A threat and risk assessment encompasses a comprehensive examination of both the users and the digital identity system for potential threats and the subsequent evaluation of the associated security risks. This assessment takes into account the likelihood and potential impact of a threat event occurring, as well as the capability of a threat actor to exploit any weaknesses within the system. Based on the level of threat and risk identified, appropriate risk management strategies can be developed, which may include the acceptance of the risk, the implementation of mitigation measures, or the adoption of avoidance strategies \cite{Shostack2014, Sheik2022}.
\vskip1pc
We expect that this combination of policy perspective, multidisciplinary academic perspectives, the perspective of end-users and stakeholders, and the technical and security perspectives will complement one another to provide a more complete picture of what is required for the development of an EC in a UK context. This, in turn, can be very useful input for policy making, regulation and inform best practices and specifications for the DIATF.
\vskip1pc
Two types of results are expected from these research activities. First, this research approach will provide a set of requirements and specifications that can be used for the design and operation of a model EC. These will take the form of commonly used artefacts that are used for this purpose, such as use case descriptions, use case diagrams, data flow diagrams and process flow diagrams. These will not be meant to provide an exhaustive set of rigid specifications, but instead will focus on the DI-specific characteristics of the design and operation of an EC. The focus will also be on addressing ‘pain points’ or ‘critical incidents’ \cite{Butterfield2005} identified in the onboarding process. These artefacts also have the potential to be used as ‘boundary objects’ to facilitate communication, engagement and feedback from stakeholders \cite{Lee2007, Vines2013}. Second, we expect to provide a set of qualitative recommendations that arise from these research activities. These recommendations will ensure that aspects of citizen inclusion and empowerment are adequately addressed (e.g., taking into account the needs of diverse groups of citizens), and will provide more flexibility in the output.
\vskip1pc
\section{Conclusion}
\vskip1pc
The implementation of a social inclusive and technically robust onboarding process for digital identities is a under-emphasised but highly impactful component of the development of digital identifies. It is an important element of the future economic success of the UK, the trust and participation of all elements of the British society in this digital future, and the strength of digitally platformed or cyber-influenced social relationships within the UK and outside. The multidisciplinary research approach outlined in this work will provide impact-laden research that can be utilised by government policy officials and technology partners to improve their DI offers.

\section*{Acknowledgements}
This research was part-funded by SPRITE+: The Security, Privacy, Identity, and Trust Engagement NetworkPlus (EPSRC grant number EP/S035869/1).

\bibliographystyle{IEEEtran}
{\small
\bibliography{refs}}

\end{multicols}

\end{document}